\documentclass[11pt]{article}
\usepackage{amssymb}

\usepackage{amsmath}
\usepackage{graphicx}
\usepackage{fullpage,amsmath}

\begin{document}

\title{Perfect fluid LRS Bianchi I with time varying constants}
\author{J.A. Belinch\'{o}n\footnote{E-mail:abelcal@ciccp.es}\\
\footnotesize{Dept. F\'{\i}sica ETS Arquitectura. 
Av. Juan de Herrera 4. Madrid 28040. Espa\~{n}a.}}
\maketitle

\begin{abstract}
It is investigated the behaviour of the ``constants'' $G,$ $c$ and $\Lambda $
in the framework of a perfect fluid LRS Bianchi I cosmological model. It has
been taken into account the effects of a $c-$variable into the curvature
tensor. Two exact cosmological solutions are investigated, arriving to the
conclusion that if $q<0$ (deceleration parameter) then $G,$ $c$ are growing
functions on time $t$ while $\Lambda $ is a negative decreasing function on
time.
\end{abstract}

\section{Introduction}

In a recent paper (see \cite{Tony2}) we have investigated the behaviour of the ``constants'' $G,
$ $c$ and $\Lambda $ in the frame work of a bulk viscous LRS Bianchi I
cosmological model where the effects of a $c-$variable into the curvature
tensor were taken into account. The main conclusion of such work is that $G$
and $c$, under the physical restrictions (thermodynamics constrains), are
growing functions on time while $\Lambda $ is a negative decreasing
function. In that paper, all the physical quantities depend of the viscous
parameter and the thermodynamical restrictions bring us to determine all the
parameters without any ambiguity, concluding that only the physical
solutions are those which equation of state is $\omega =1$ (ultra stiff
matter) and $G$ and $c$ rea growing functions on time while $\Lambda $
behaves as a negative decreasing function.

As we have showed in such paper for late time viscous, effects vanish in
such a way that the model tends to a perfect fluid era. In this paper we try
to study precisely this situation. Taking into account the effects of a $c-$%
variable into the curvature tensor we outline the field equations for a
perfect fluid cosmological model with LRS Bianchi I symmetries. Under the
assumed hypotheses we shall study two exact solutions. In this occasion we
will not be able to determine the numerical values of the constants without
any ambiguity in such a way that only imposing the condition $q<0$ (i.e.
that the model accelerates) we arrive to the conclusion that $G$ and $c$ are
growing while $\Lambda $ is negative and decreasing. We believe that these
solutions must be correct since in some way these solutions have to connect
with the viscous one. 

The paper is organized as follows. In section 2 we outline the field
equations and the assumed hypotheses in order to find the exact solutions.
In section 3 we show the solution of the two studied models. 

In the first of them, the solution has been obtained under the hypotheses
that the scale factor are follow the relationship $X(t)\thickapprox Y(t)^{n},
$ this hypothesis is standard in this class of models. We also assumed that the
speed of light, $c(t),$ follows a power law dependence on time $t,$ i.e. $%
c(t)=c_{0}t^{a},$ where $a$ is a undetermined numerical constants. With
these two hypotheses it is founded a singular exact solution to the field
equations, but as we have mentioned above, in this case, we are only able to
find poor restrictions to the numerical values. In order to see how
each quantity behaves, we have plotted some cases, finding that the ``constants''  $G
$ and $c$ can be growing as well as decreasing \ functions (with some
restrictions) while $\Lambda $ is a negative decreasing function on time $t.$
In the same way we show that it is possible to find (fine tuning) cases
which verify the relationship $G/c^{2}=const..$ Only imposing some
restrictions, like $q<0,$ it is founded that $G$ and $c$ must be growing
while $\Lambda $ behaves as a negative decreasing function.

In the second studied solution founded under the hypotheses $%
c(t)\thickapprox Y(t)^{n}$ and $X(t)\thickapprox Y(t)^{m}$, we shall show
that it is non-singular. In the same way than in the first solution we shall
find some restrictions to our model and we will plot some cases in order to
see how the main quantities behave. As in the above solution we arrive to
the conclusion that the most probable solution is those where $G$ and $c$
are growing functions while $\Lambda $ is a decreasing function.

\section{The model and the hypotheses}

We consider a locally rotationally symmetric (LRS) Bianchi I (\cite{Ha}-\cite
{Cha}) spacetime with metric 
\begin{equation}
ds^{2}=-c^{2}(t)dt^{2}+X^{2}(t)dx^{2}+Y^{2}(t)\left( dy^{2}+dz^{2}\right) ,
\label{line}
\end{equation}
where the gravitational field equations with variable $G$, $c$ and $\Lambda $
are: 
\begin{equation}
R_{ik}-\frac{1}{2}g_{ik}R=\frac{8\pi G(t)}{c^{4}\left( t\right) }%
T_{ik}+\Lambda (t)g_{ik},  \label{ECU1}
\end{equation}
where the energy momentum tensor is: 
\begin{equation}
T_{ij}=\left( \rho +p\right) u_{i}u_{j}-pg_{ij},  \label{s2-e2}
\end{equation}
and where $p$ and $\rho $ satisfy the usual equation of state, $p=\omega
\rho $ in such a way that $\omega \in \left( -1,1\right] ,$ that is to say,
our universe is modeled by a perfect fluid$.$

Applying the covariance divergence to the second member of equation (\ref
{ECU1}) we get: 
\begin{equation}
div\left( \frac{G}{c^{4}}T_{i}^{j}+\delta _{i}^{j}\Lambda \right) =0,
\label{conser1}
\end{equation}
hence 
\begin{equation}
T_{i;j}^{j}=\left( \frac{4c_{,j}}{c}-\frac{G_{,j}}{G}\right) T_{i}^{j}-\frac{%
c^{4}\delta _{i}^{j}\Lambda _{,j}}{8\pi G},  \label{conser2}
\end{equation}
that simplifies to: 
\begin{equation}
\dot{\rho}+\left( \rho +p\right) H=-\frac{\dot{\Lambda}c^{4}}{8\pi G}-\rho 
\frac{\dot{G}}{G}+4\rho \frac{\dot{c}}{c},  \label{conser3}
\end{equation}

Therefore, with all these assumptions and taking into account the
conservation principle, i.e., $div(T_{i}^{j})=0$, the resulting field
equations are as follows:

\begin{align}
2\frac{\dot{X}}{X}\frac{\dot{Y}}{Y}+\frac{\dot{Y}^{2}}{Y^{2}}& =\frac{8\pi G%
}{c^{2}}\rho +\Lambda c^{2},  \label{nfield1} \\
2\frac{\ddot{Y}}{Y}+\frac{\dot{Y}^{2}}{Y^{2}}-2\frac{\dot{Y}}{Y}\frac{\dot{c}%
}{c}& =-\frac{8\pi G}{c^{2}}\left( p\right) +\Lambda c^{2},  \label{nfield2}
\\
\frac{\ddot{Y}}{Y}+\frac{\dot{X}}{X}\frac{\dot{Y}}{Y}+\frac{\ddot{X}}{X}%
-\left( \frac{\dot{X}}{X}+\frac{\dot{Y}}{Y}\right) \frac{\dot{c}}{c}& =-%
\frac{8\pi G}{c^{2}}\left( p\right) +\Lambda c^{2},  \label{nfield22} \\
\dot{\rho}+\left( \omega +1\right) \rho \left( \frac{\dot{X}}{X}+2\frac{\dot{%
Y}}{Y}\right) & =0,  \label{nfield3} \\
\frac{\dot{\Lambda}c^{4}}{8\pi G}+\rho \frac{\dot{G}}{G}-4\rho \frac{\dot{c}%
}{c}& =0.  \label{nfield5}
\end{align}

Since we have defined the $4-$velocity as: 
\begin{equation}
u^{i}=\left( \frac{1}{c(t)},0,0,0\right) \text{ \ \ \ \ \ \ \ / \ \ \ \ \ \
\ }u^{i}u_{j}=-1,
\end{equation}
then the expansion is defined as follows 
\begin{equation}
\theta :=u_{\,;\,i}^{i}\text{, \ \ \ \ \ \ \ \ \ }\theta =\frac{1}{c(t)}%
\left( \frac{\dot{X}}{X}+2\frac{\dot{Y}}{Y}\right),  \label{expan}
\end{equation}
and the shear is 
\begin{equation}
\sigma ^{2}=\frac{1}{2}\sigma _{ij}\sigma ^{ij},\text{ \ \ \ \ \ \ }\sigma =%
\frac{\sqrt{3}}{3c(t)}\left( \frac{\dot{X}}{X}-\frac{\dot{Y}}{Y}\right).
\label{shear}
\end{equation}

Curvature is described by the tensor field $R_{jkl}^{i}.$ It is well know
that if one uses the singular behaviour of the tensor components \ or its
derivates as a criterion for singularities, one gets into trouble since the
singular behaviour of the coordinates or the tetrad basis rather than the
curvature tensor. To avoid this problem, one should examine the scalars
formed out of the curvature. The invariants $K_{1}$ and $K_{2}$ (the
Kretschmann scalars) are very useful for the study of the singular
behaviour: 
\begin{eqnarray}
K_{1} &:&=R_{ijkl}R^{ijkl}=\frac{4}{c^{4}}\left[ \left( \frac{\ddot{X}}{X}%
\right) ^{2}-2\frac{\ddot{X}}{X}^{2}\frac{\dot{c}\dot{X}}{cX}+\frac{\dot{c}%
^{2}\dot{X}^{2}}{c^{2}X^{2}}+2\left( \frac{\ddot{Y}}{Y}\right) ^{2}-\right.
\nonumber \\ 
&&\left. -4\frac{\ddot{Y}}{Y}\frac{\dot{Y}}{Y}\frac{\dot{c}}{c}+\left( \frac{%
\dot{Y}\dot{c}}{Yc}\right) ^{2}+2\left( \frac{\dot{X}}{X}\frac{\dot{Y}}{Y}%
\right) ^{2}+\left( \frac{\dot{Y}}{Y}\right) ^{4}\right] , \label{k1}
\end{eqnarray}
\begin{eqnarray}
K_{2} &:&=R_{ij}R^{ij}=\frac{2}{c^{4}}\left[ \left( \frac{\ddot{X}}{X}%
\right) ^{2}+\left( \frac{\dot{c}\dot{X}}{cX}\right) ^{2}-2\frac{\ddot{X}}{X}%
^{2}\frac{\dot{c}\dot{X}}{cX}+3\left( \frac{\dot{X}}{X}\frac{\dot{Y}}{Y}%
\right) ^{2}+\left( \frac{\dot{Y}}{Y}\right) ^{4}+3\left( \frac{\ddot{Y}}{Y}%
\right) ^{2}\right. \nonumber \\ 
&&+3\left( \frac{\dot{Y}}{Y}\frac{\dot{c}}{c}\right) ^{2}-6\frac{\ddot{Y}}{Y}%
\frac{\dot{Y}}{Y}\frac{\dot{c}}{c}+2\frac{\ddot{X}}{X}\frac{\ddot{Y}}{Y}-2%
\frac{\ddot{X}}{X}\frac{\dot{Y}}{Y}\frac{\dot{c}}{c}-2\frac{\ddot{Y}}{Y}%
\frac{\dot{c}\dot{X}}{cX}+2\frac{\dot{X}}{X}\frac{\dot{Y}}{Y}\left( \frac{%
\dot{c}}{c}\right) ^{2}+ \nonumber\\ 
&&+2\frac{\dot{X}}{X}\frac{\dot{Y}}{Y}\frac{\ddot{X}}{X}-2\left( \frac{\dot{X%
}}{X}\right) ^{2}\frac{\dot{Y}}{Y}\frac{\dot{c}}{c}+2\frac{\ddot{Y}}{Y}%
\left( \frac{\dot{Y}}{Y}\right) ^{2}+2\frac{\ddot{Y}}{Y}\frac{\dot{Y}}{Y}%
\frac{\dot{X}}{X}-2\frac{\dot{X}}{X}\left( \frac{\dot{Y}}{Y}\right) ^{2}%
\frac{\dot{c}}{c}- \nonumber\\ 
&&\left. -2\left( \frac{\dot{Y}}{Y}\right) ^{3}\frac{\dot{c}}{c}+2\left( 
\frac{\dot{Y}}{Y}\right) ^{3}\frac{\dot{X}}{X}\right].  \label{K2}
\end{eqnarray}

\subsection{Simplifying hypotheses}

In order to solve these equations we shall need to make the following
simplifying hypotheses:

\begin{enumerate}
\item  From the equations (\ref{nfield2}) and (\ref{nfield22}) it is
obtained 
\begin{equation}
\frac{\ddot{Y}}{Y}+\frac{\dot{Y}^{2}}{Y^{2}}+\left( \frac{\dot{X}}{X}-\frac{%
\dot{Y}}{Y}\right) \frac{\dot{c}}{c}-\frac{\dot{X}}{X}\frac{\dot{Y}}{Y}-%
\frac{\ddot{X}}{X}=0 . \label{M_T}
\end{equation}

\begin{enumerate}
\item  Then we can make the next hypothesis: we impose that
\begin{equation}
X\thickapprox Y^{n},
\end{equation}
therefore equation (\ref{M_T}) yields 
\begin{equation}
\ddot{Y}=\left( \frac{n^{2}-1}{1-n}\right) \frac{\dot{Y}^{2}}{Y}+\dot{Y}%
\frac{\dot{c}}{c},
\end{equation}
obtaining a trivial solution 
\begin{equation}
Y=\left( C_{2}(n+2)+C_{1}(n+2)\int c(t)dt\right) ^{\frac{1}{(n+2)}},
\end{equation}
then we will need to make a hypothesis about the behaviour of $c(t)$ in
order to obtain a complete solution.

\item  If we make the next assumption 
\begin{equation}
c(t)=c_{0}Y^{n},
\end{equation}
then eq. (\ref{M_T}) has the following solution: 
\begin{equation}
Y=X\left( C_{2}\left( n-2\right) \int X^{n-3}dt+C_{1}(2-n)\right) ^{\frac{1}{%
2-n}},
\end{equation}
then we will need to make a hypothesis about the behaviour of $X(t)$ in
order to obtain a complete solution.
\end{enumerate}
\end{enumerate}

In the next section we shall study two different solutions which correspond
to each one of the hypotheses.

\section{Solutions}

\subsection{Model 1. Hypotheses 1a.}

If we follow the hypotheses $1a$ then it is observed that 
\begin{equation}
Y=\left( C_{2}(n+2)+C_{1}(n+2)\int c(t)dt\right) ^{\frac{1}{(n+2)}},
\label{bea}
\end{equation}
and making the following hypothesis about the behaviour of $c(t)$ 
\begin{equation}
c(t)=c_{0}t^{a},  \label{sol1-c}
\end{equation}
with $a\in \mathbb{R},$ which is possibly the most simple hypothesis about
it behaviour. We have choice this form taken into account the results
obtained in our previous paper (see \cite{Tony1}) where we have founded,
through the Lie analysis of the differential equations, that the only
physical solution for the function $c(t)$ follows a power-law dependence of
time $t.$ We can choice other form for this function and with the help of
eq. (\ref{bea}) it is obtained the exact form of the scale factors $Y(t)$
and $X(t),$ which brings us to a very different scenario$.$ Therefore, for
our election of $c(t)$ (\ref{sol1-c}) we found that $Y(t)$ and $X(t)$
behaves as 
\begin{equation}
Y=\left( C_{2}(n+2)+C_{1}\frac{(n+2)}{a+1}t^{a+1}\right) ^{\frac{1}{(n+2)}},%
\text{ \ \ \ \ }\Longrightarrow \text{\ \ \ }X=\left( C_{2}(n+2)+C_{1}\frac{%
(n+2)}{a+1}t^{a+1}\right) ^{\frac{n}{(n+2)}},
\end{equation}
observing that we must to impose the following restrictions on the possible
values of the constants $a$ and $n,$ in this way, we see that $a\neq -1$ and 
$n>-2$ if we want that our scale factor $Y(t)$ be a growing function.

Now defining 
\begin{equation}
H=\left( \frac{\dot{X}}{X}+2\frac{\dot{Y}}{Y}\right) =\left( 2+n\right) 
\frac{\dot{Y}}{Y}=\frac{\left( a+1\right) C_{1}t^{a}}{\left( C_{2}\left(
a+1\right) +C_{1}t^{a+1}\right) },
\end{equation}
we can calculate the deceleration parameter $q$%
\begin{equation}
q=\frac{d}{dt}\left( \frac{3}{H}\right) -1=3\left( -a\frac{C_{2}}{C_{1}}%
t^{-a-1}+\frac{1}{a+1}\right) -1,
\end{equation}
as we can see this expression only has sense if $C_{2}=0,$ therefore 
\begin{equation}
q=\frac{2-a}{a+1},  \label{q2a}
\end{equation}
as it is observed we only have $q<0$ if $a>2.$

Now taking into account the expression 
\begin{equation}
\dot{\rho}+\left( \rho +p\right) H=0,\Longrightarrow \rho
=d_{0}Y^{-(n+2)\left( \omega +1\right) },
\end{equation}
hence the energy density behaves as 
\begin{equation}
\rho =d_{0}t^{-\left( \omega +1\right) },
\end{equation}
being $d_{0}=const.$

Now from eq. 
\begin{equation}
\left( 2n+1\right) \frac{\dot{Y}^{2}}{Y^{2}}=\frac{8\pi G}{c^{2}}\rho
+\Lambda c^{2},
\end{equation}
and defining 
\begin{equation}
\left( 2n+1\right) \frac{\dot{Y}^{2}}{Y^{2}}=f(t),
\end{equation}
we obtain $\Lambda $ as 
\begin{equation}
\Lambda =\left( f(t)-\frac{8\pi G}{c^{2}}\rho (t)\right) \frac{1}{c^{2}(t)},
\label{lam1}
\end{equation}
and substituting this expression into 
\begin{equation}
\frac{\dot{\Lambda}c^{4}}{8\pi G\rho }+\frac{\dot{G}}{G}-4\frac{\dot{c}}{c}%
=0,
\end{equation}
we end obtaining the behaviour of the ``constant'' $G$ as: 
\begin{equation}
G=\frac{c^{2}f}{8\pi \dot{\rho}}\left( \frac{\dot{f}}{f}-2\frac{\dot{c}}{c}%
\right) ,
\end{equation}
therefore it yields 
\begin{equation}
G=G_{0}t^{3a+\omega (a+1)-1},  \label{G2a}
\end{equation}
being $G_{0}=const..$ As it is observed, the quotient between $G$ and $c$
behaves as 
\begin{equation}
\frac{G}{c^{2}}\thickapprox t^{a+\omega (a+1)-1},
\end{equation}
finding that we only have the relationship 
\begin{equation}
\frac{G}{c^{2}}=const.\Longleftrightarrow a=\frac{1-\omega }{(1+\omega )}.
\label{gc21}
\end{equation}
We insist into emphasize this relationship, because in (\cite{Tony1}) we
obtain this expression as integration condition (with out any supplementary
hypothesis) i.e. at least this expression could have mathematical sense (as
symmetry condition). In a subsequent paper (\cite{Tony2}), where we were
studying a LRS Bianchi I viscous model, this relationship was also verifies
iff the viscous parameter was the usual one i.e. $\gamma =1/2$. In the next
subsection (see below) this relationship help us to determine the values of
the numerical constants.

Once it has been obtained $G$ then we back to eq. (\ref{lam1}) obtaining in
this way the behaviour of $\Lambda :$%
\begin{equation}
\Lambda =\Lambda _{0}t^{-2(a+1)},  \label{L2a}
\end{equation}
where $\Lambda _{0}=const.$

The behaviour of the Krestchmann scalars are: 
\begin{equation}
K_{1}\thickapprox t^{-4(a+1)},\text{ \ \ \ \ \ \ \ \ \ \ \ }%
K_{2}\thickapprox t^{-4(a+1)},
\end{equation}
while the expansion and the shear behaves as: 
\begin{equation}
\theta =\frac{a+1}{c_{0}}t^{-(a+1)},
\end{equation}
\begin{equation}
\sigma =\frac{\sqrt{3}}{3}\frac{\left( n-1\right) }{\left( n+2\right) }%
\left( a+1\right) t^{-(a+1)}.
\end{equation}

It is observed that all the quantities depend only of the constants $a$ and $%
\omega $ and only $X(t)$ depends of $n.$ Therefore from these results we
only can say that $n>-2,$ $\omega >-1$ and $a>-1,$ we have not more
restrictions for these constants except that if we want that our solution
verifies the condition $q<0$ then $a>2,$ but this condition tells us that $%
c(t)$ is a growing function on time $t.$

\subsubsection{Conclusion for this model. Numerical values and graphics for
the main quantities.}

In the viscous case discussed in (\cite{Tony2}) we were able to find a
concrete physical values for the constants $\left( a,n,\omega \right) $
under thermodynamical considerations but in this case we only have a poor
restrictions, $n>-2,$ $\omega >-1$ and $a>-1$ but if $q<0$ then $a>2.$ But
we cannot to discard the case $a=-1/2$, this possibility is so valid than
the case $a=1/2.$ In this subsection we will study numerically some of the
possible cases in order to show the behaviour of the different quantities.
In the solutions exposed above we have written, for example $G_{0}$ or $%
\Lambda _{0}$ as constants, but these numerical constants are horrible
expression that depend of the constants $\left( a,n,\omega \right) $ and
other numerical factors. The next figures are sensible to all these
numerical factors. We shall study five cases, for this purpose we need to
fix some numerical constants $\left( a,n\right) $ as well as consider
different equations of state $\left( \omega \right) $. Since we only have
the above restrictions (which do not help us) we can choose these
numerical values in a very arbitrary way but we prefer to ``assume'' that eq. (\ref
{gc21}) is verified (since in previous works this relationship was obtained
as integration condition and hence we believe that our model, and without
other restrictions, must verifies it) in this way we find that the constant $%
a=\psi (\omega )$ i.e. is a function of the equation of state and to fix
arbitrarily the value of $n$ since it only affects directly to the scale factor 
$X(t).$

In the rest of this section we shall consider the following values for the
numerical constants 
\begin{equation*}
C_{1}=1,\quad C_{2}=0,\quad c_{0}=1,\quad d_{0}=1,
\end{equation*}
and in the following table is summarized the colour of each solution which
correspond to the different values of $\left( a,n,\omega \right) $ and the
corresponding value of the parameter $q:$ 
\begin{equation*}
\begin{array}{|c|c|c|c|}
colour & \omega  & a & n \\ \hline
red & 1 & -1/3 & 3/2 \\  \hline
blue & 1/3 & 1/2 & 1/2 \\  \hline
magenta & 0 & 1 & 1/5 \\  \hline
black & -1/3 & 2 & 5/2 \\  \hline
points & -2/3 & 5 & 2/3
\end{array}
\begin{array}{||c|}
q \\  \hline
7/2>0 \\  \hline
1>0 \\  \hline
1/2>0 \\  \hline
0 \\  \hline
-1/2<0
\end{array}
\end{equation*}

We would like to emphasize that the red colour solution i.e. it has been
calculated with the following numerical values $\left( \omega
=1,a=-1/3,n=3/2\right) $ does not verifies the condition (\ref{gc21}) since
for $\omega =1,a=0$ and it has $q>0,$ the rest of the solutions verify such
relationship. The only solution that verifies the condition $q<0$ corresponds
to the equation of state $\omega =-2/3$ which is in agreement with the
recent observations and theoretical models. 

\begin{figure}[h!]
\begin{center}
\includegraphics[height=2.194in,width=2.194in]{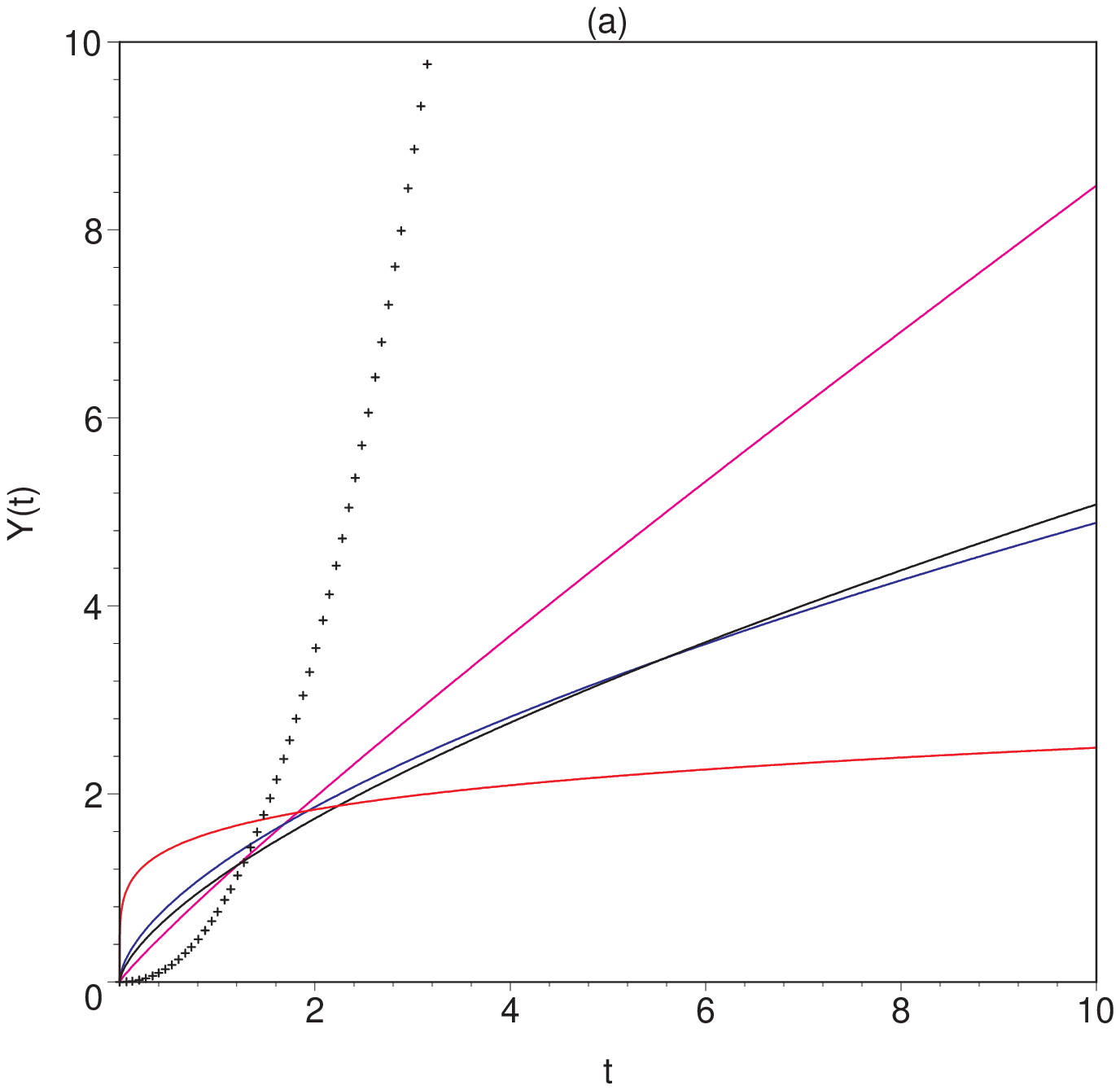} %
\includegraphics[height=2.194in,width=2.194in]{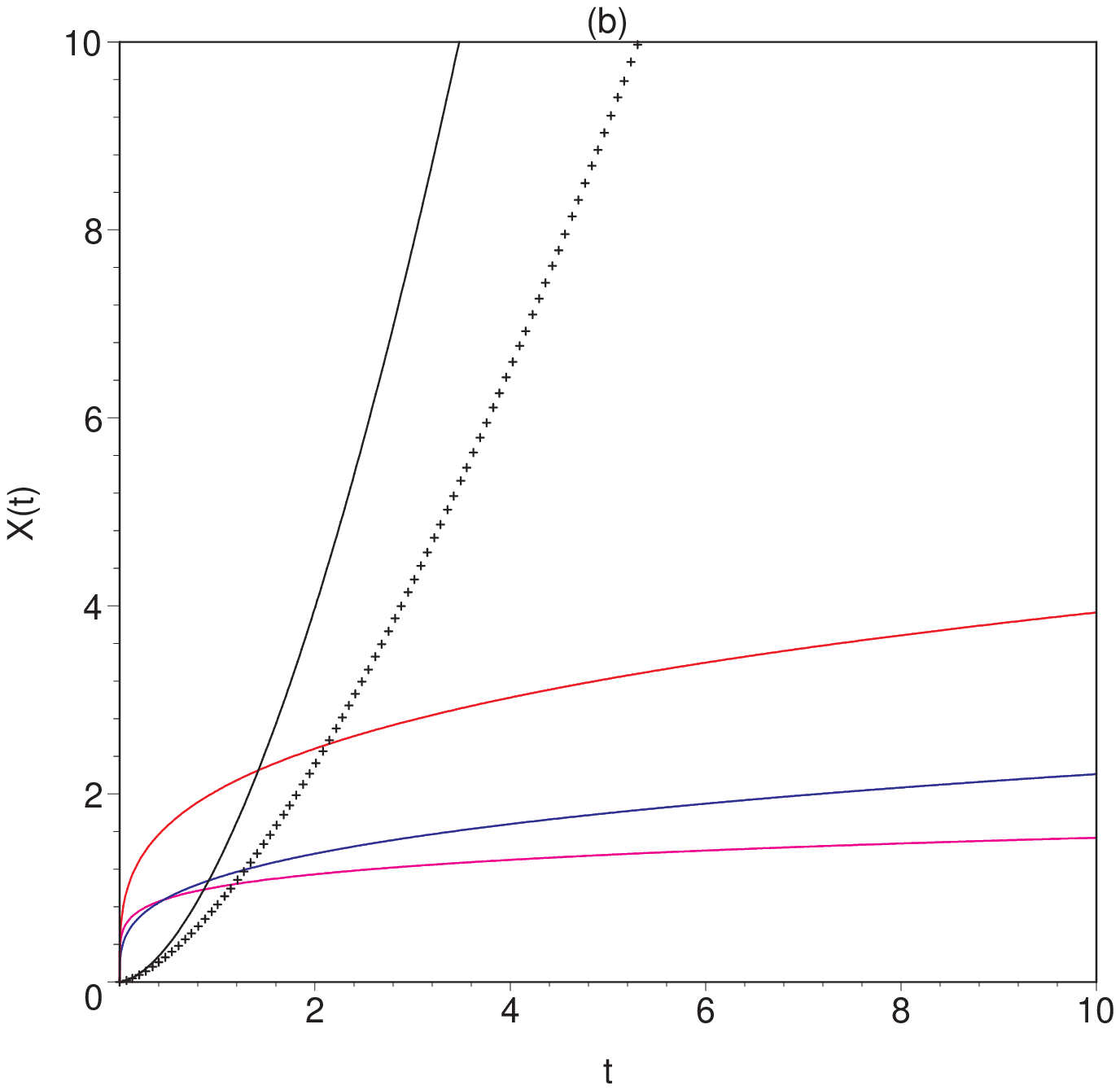}
\end{center}
\caption{In the figure (a) it is plotted the radius $Y(t)$ while in figure
(b) it is plotted the variation of the radius $X(t)$. Both radius have a
singular start and are growing functions on time.}
\label{Radios2}
\end{figure}

In the first place we begin studying the behaviour of the radius of our
model. With these numerical values we can see in fig. (\ref{Radios2}) the
different behaviour of our scale factors. In all the studied cases we can
see that these scale factors are growing functions on time $t$. These
solutions are singular as the Krestchmann invariants show us. 

\begin{figure}[h!]
\begin{center}
\includegraphics[height=2.194in,width=2.194in]{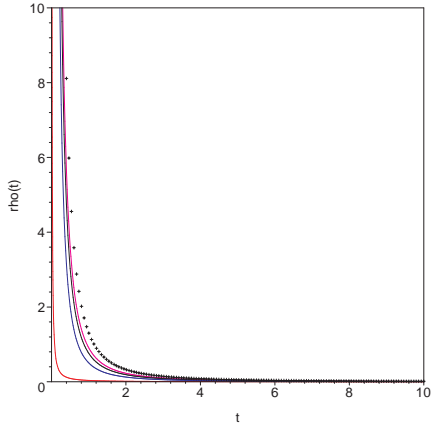}
\end{center}
\caption{The figure shows the variation of the energy density $\protect\rho%
(t)$.}
\label{densidad2}
\end{figure}

With regard to the energy density (see fig. (\ref{densidad2})) we can see
that all the solutions have physical meaning since this quantity depends
only of the equation of state.

\begin{figure}[h!]
\begin{center}
\includegraphics[height=2.194in,width=2.194in]{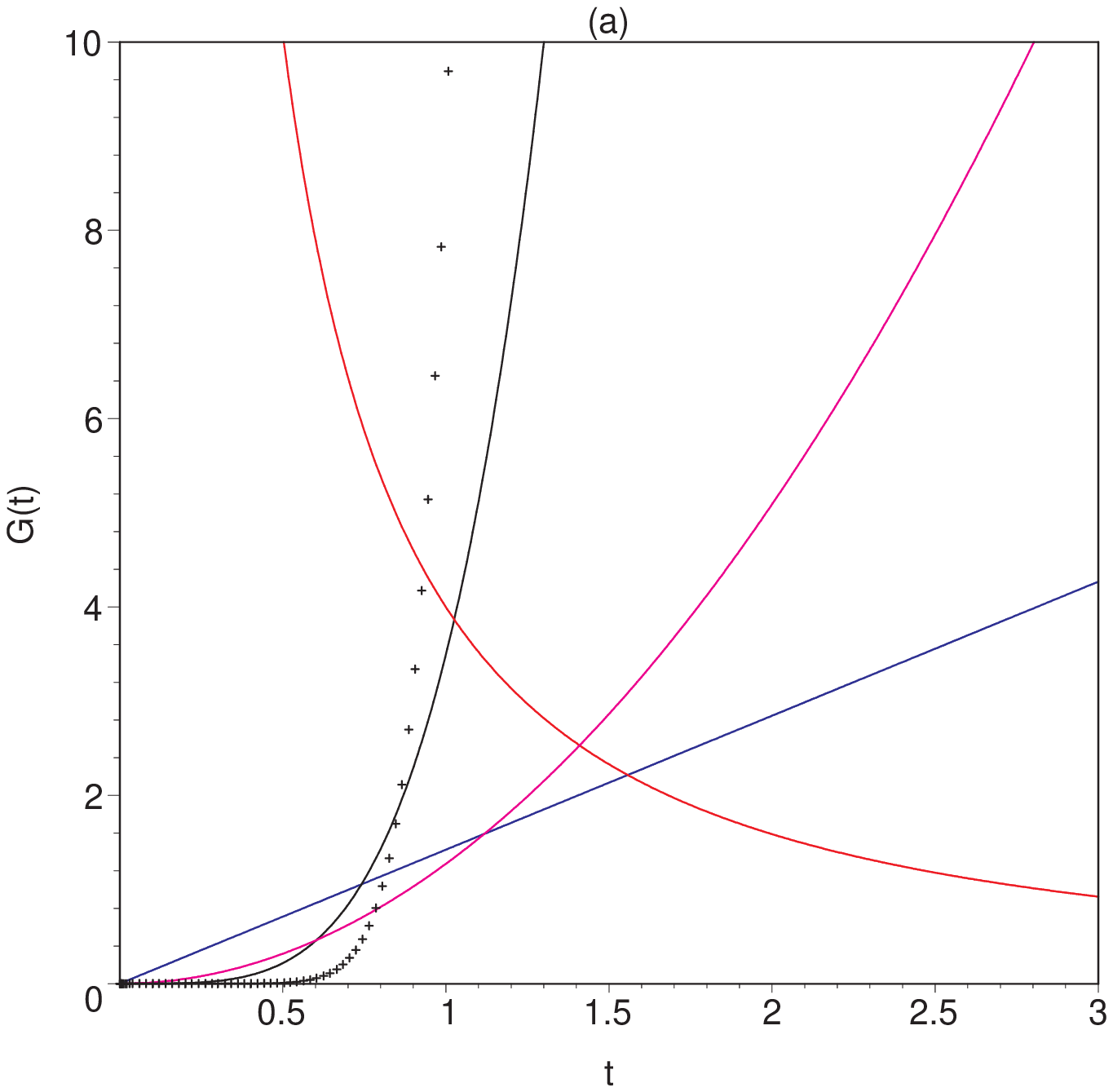} %
\includegraphics[height=2.194in,width=2.194in]{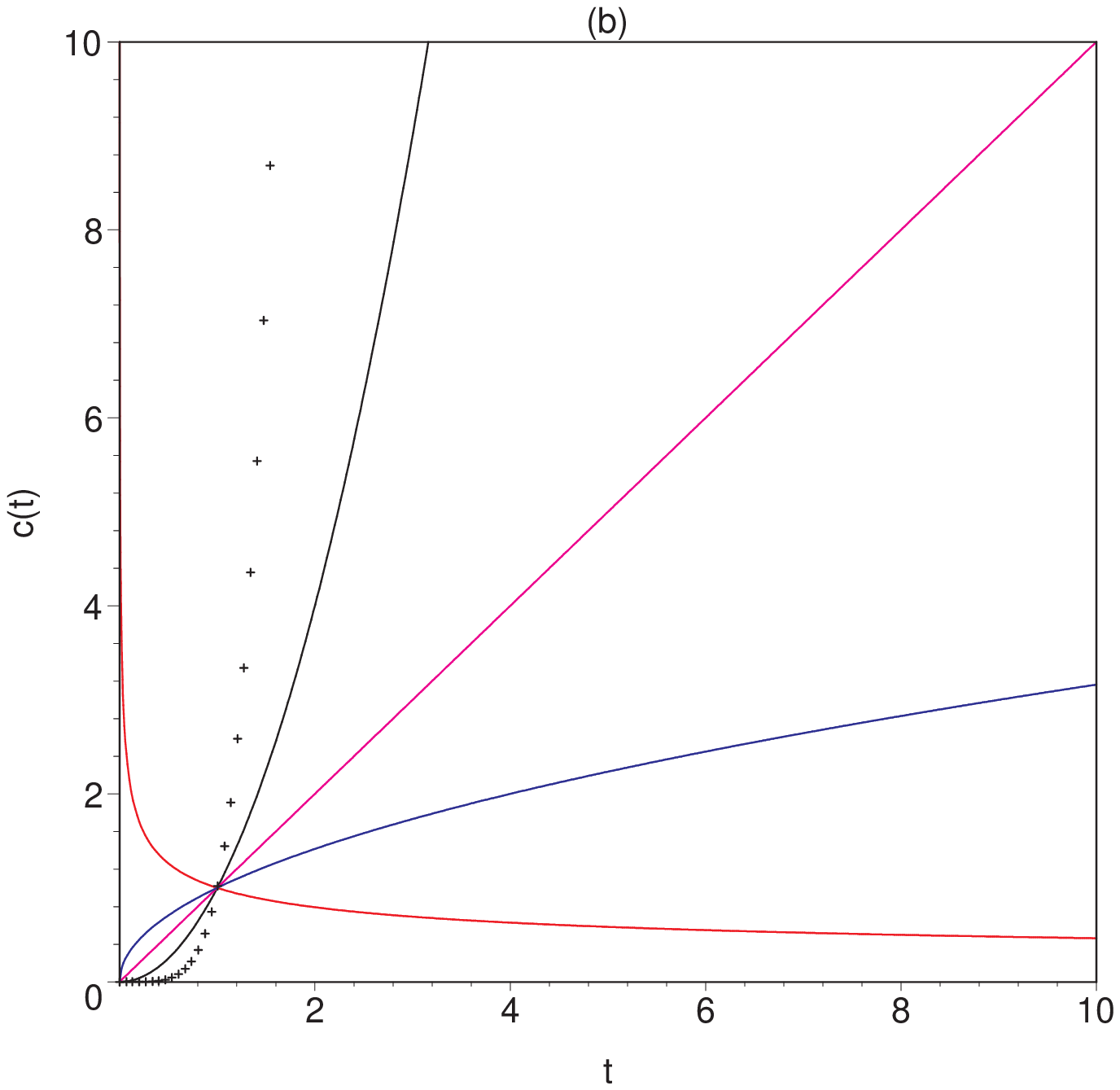} %
\includegraphics[height=2.194in,width=2.194in]{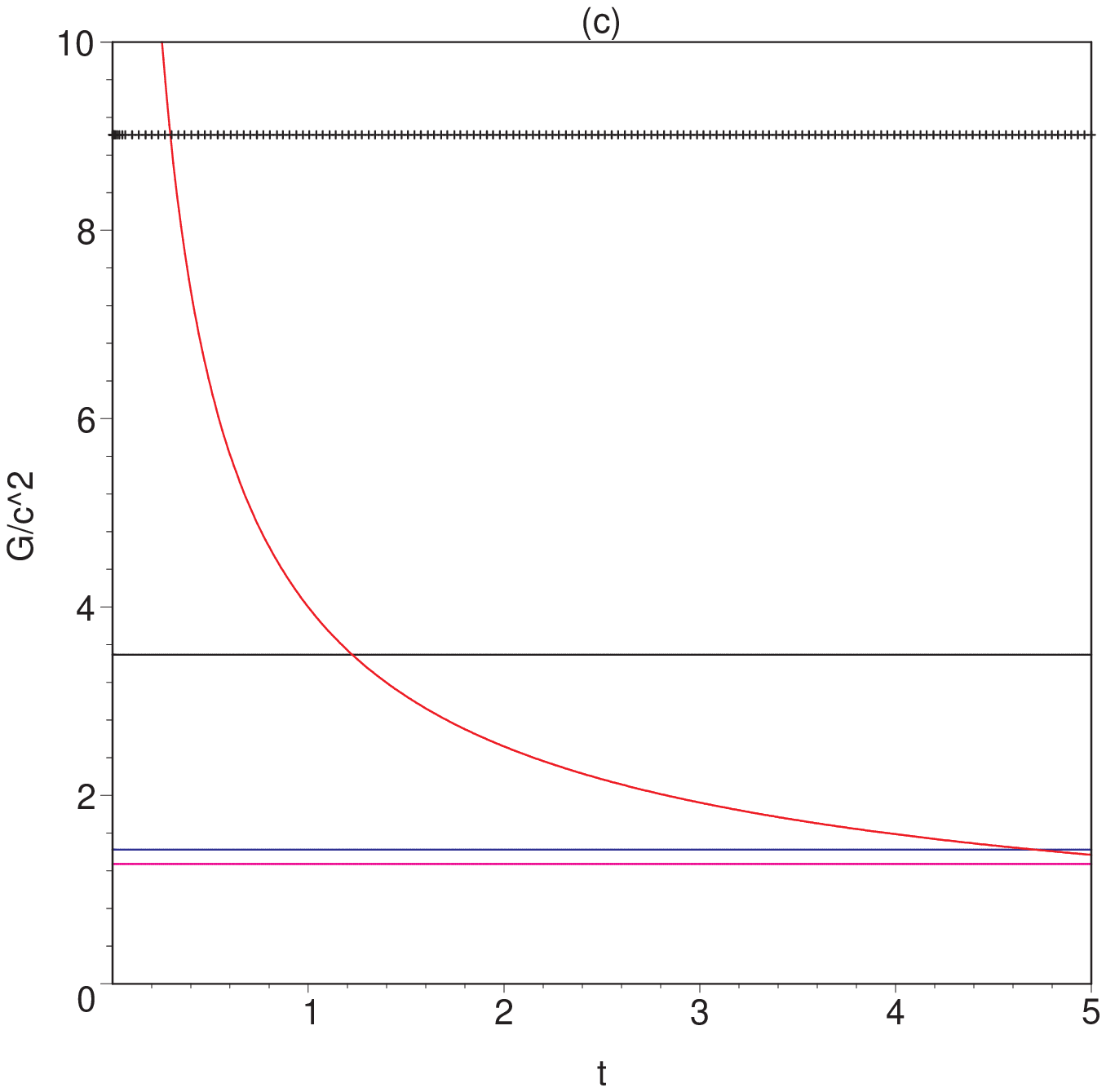}
\end{center}
\caption{In the figure (a) we show the variation of ``constant" $G(t)$. In
the figure (b) we show the variation of the ``constant" $c(t)$ and in the
figure (c) it is plotted the relationship $G/c^{2}$.}
\label{constantes2}
\end{figure}

The variation of the ``constants'' $G$ and $c$ as well as the relationship $%
G/c^{2}$ is shown in fig. (\ref{constantes2}). These figures show us that
both ``constants'' are growing functions on time $t$ except in the case (red
line) $a=-1/3$ where $G$ and $c$ are decreasing functions on time $t,$ and
as we have pointed out above this solution does not verify the condition $%
G/c^{2}=const.$. The rest of solutions verify the relationship $%
G/c^{2}=const.$ since we have chosen the numerical values with such purpose.

\begin{figure}[h!]
\begin{center}
\includegraphics[height=2.194in,width=2.194in]{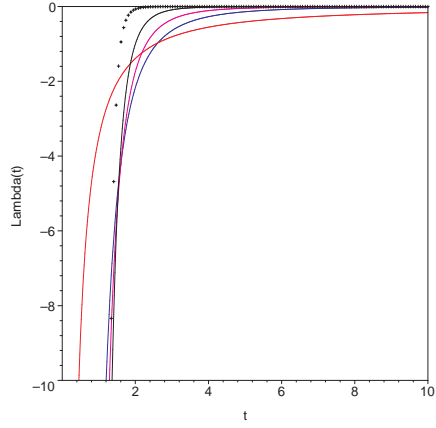}
\end{center}
\caption{The variation of the cosmological ``constant" $\Lambda(t)$.}
\label{lambda2}
\end{figure}

With regard to the cosmological ``constant'', see fig. (\ref{lambda2}), it
is observed that all the solutions are decreasing but negative. Maybe this
fact tells us that we may consider $\Lambda (t)$ as a true ghost energy
density.

\begin{figure}[h!]
\begin{center}
\includegraphics[height=2.194in,width=2.194in]{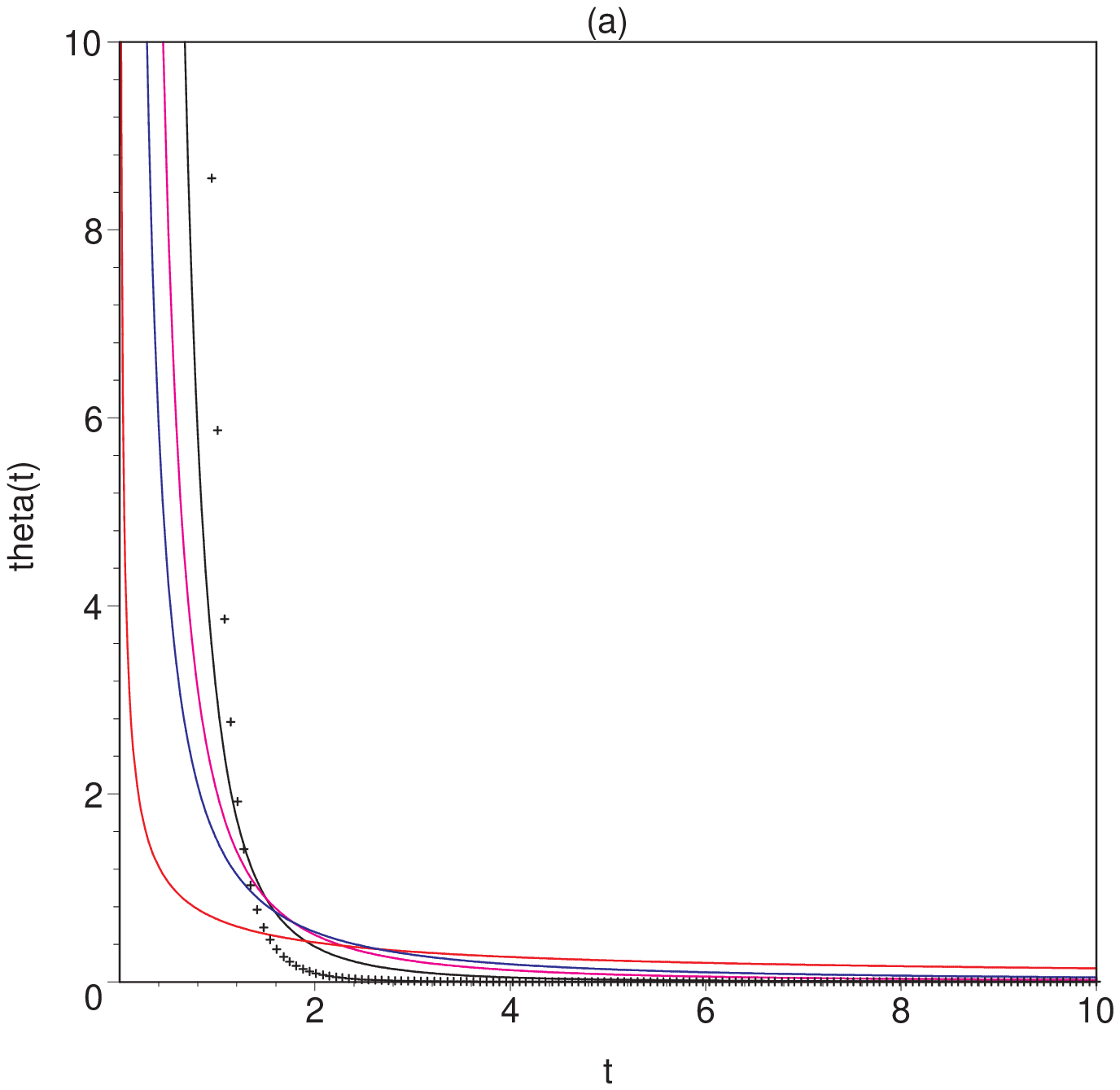} %
\includegraphics[height=2.194in,width=2.194in]{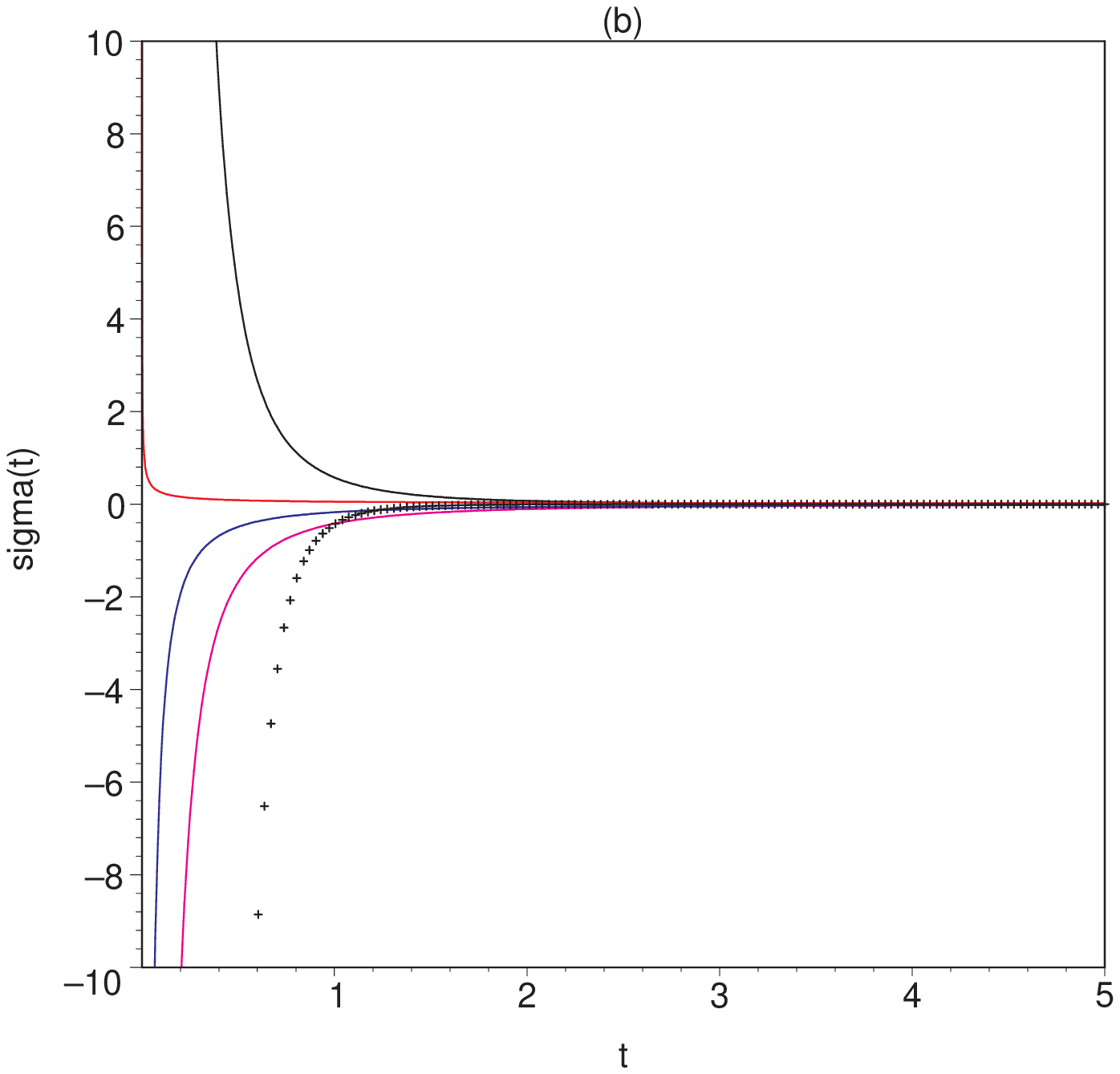}
\end{center}
\caption{In figure (a) it is plotted the expansion $\protect\theta (t)$,
while in figure (b) it is plotted the shear $\protect\sigma (t)$.}
\label{expansion2}
\end{figure}

The expansion and the shear behave as follows, see fig. (\ref{expansion2}).
As we can see all the models studied show a decreasing expansion and the
only models that have a positive shear are the plotted with the red and black
color lines. The shear and expansion scalars calculated above indicate that
the Universe is shearing and expanding with time. The shear, which is a
degree of anisotropy in the Universe, decreases monotonically with time.
This indicates the fact that the initially anisotropic Universe gradually
tends to an isotropic Universe at late time (present epoch) which is in
agreement with the recent observations of a negligible amount of anisotropy
present in the CMBR.

Other solutions can be studied changing the numerical values of the
constants as well as taking into account different equations of state. We
have found at least a solution in argument with the recent observations $%
\left( q<0\right) $ (black point) which equations of state are $\omega =-2/3$
(domain walls) with $G$ and $c$ as growing functions on time $t$ and
verifying $G/c^{2}=const..$

\subsection{Model 2. Hypotheses 1b.}

If we follow the hypotheses $1b$ then it is observed that if we make the
next assumption 
\begin{equation}
c(t)=c_{0}Y^{n},
\end{equation}
then eq. (\ref{M_T}) has the following solution: 
\begin{equation}
Y=X\left( C_{2}\left( n-2\right) \int X^{n-3}dt+C_{1}(2-n)\right) ^{\frac{1}{%
2-n}},
\end{equation}
then we will need to make a hypothesis about the behaviour of $X(t)$ in
order to obtain a complete solution. Now if we assume that 
\begin{equation}
X=Y^{m},
\end{equation}
with $m\neq 1,$then 
\begin{equation}
Y=\left( C_{1}\left( m-n+2\right) t+C_{2}\left( m-n+2\right) \right) ^{\frac{%
1}{\left( m-n+2\right) }},
\end{equation}
with the restriction $n<m+2.$ Defining 
\begin{equation}
H=\left( \frac{\dot{X}}{X}+2\frac{\dot{Y}}{Y}\right) =\left( m+2\right) 
\frac{\dot{Y}}{Y}=\frac{\left( m+2\right) C_{1}}{\left( m-n+2\right) \left(
C_{1}t+C_{2}\right) },
\end{equation}
we can calculate the deceleration parameter $q$%
\begin{equation}
q=\frac{d}{dt}\left( \frac{3}{H}\right) -1=2-\frac{3n}{m+2},
\end{equation}
finding that 
\begin{equation}
q<0\Longleftrightarrow 3n>2\left( m+2\right) .  \label{qc}
\end{equation}

Now taking into account the expression 
\begin{equation}
\dot{\rho}+\left( \rho +p\right) H=0,\Longrightarrow \rho =d_{0}Y^{-\left(
2+m\right) \left( \omega +1\right) },
\end{equation}
we find the behaviour of the energy density
\begin{equation}
\rho =d_{0}\left( C_{1}\left( m-n+2\right) t+C_{2}\left( m-n+2\right)
\right) ^{\frac{-\left( 2+m\right) \left( \omega +1\right) }{\left(
m-n+2\right) }},
\end{equation}
with $d_{0}=const.$

Now from 
\begin{equation}
\left( 2n+1\right) \frac{\dot{Y}^{2}}{Y^{2}}=\frac{8\pi G}{c^{2}}\rho
+\Lambda c^{2},
\end{equation}
where 
\begin{equation}
\left( 2n+1\right) \frac{\dot{Y}^{2}}{Y^{2}}=f(t),
\end{equation}
we obtain $\Lambda $ as
\begin{equation}
\left( f(t)-\frac{8\pi G}{c^{2}}\rho (t)\right) \frac{1}{c^{2}(t)}=\Lambda ,
\label{lam2}
\end{equation}
and substituting this expression into 
\begin{equation}
\frac{\dot{\Lambda}c^{4}}{8\pi G\rho }+\frac{\dot{G}}{G}-4\frac{\dot{c}}{c}%
=0,
\end{equation}
we end obtaining the behaviour of the ``constant'' $G$ as
\begin{equation}
G=\frac{c^{2}f}{8\pi \dot{\rho}}\left( \frac{\dot{f}}{f}-2\frac{\dot{c}}{c}%
\right) =G_{0}\frac{\left( C_{1}\left( m-n+2\right) t+C_{2}\left(
m-n+2\right) \right) ^{\frac{2n+4+m+\omega \left( n+2\right) }{\left(
m-n+2\right) }}}{(C_{1}t+C_{2})^{3}},
\end{equation}
with $G_{0}=const..$ As we can see, in this occasion, the relationship
between the quantities $G$ and $c$ behaves as: 
\begin{equation}
\frac{G}{c^{2}}\thickapprox \frac{\left( C_{1}\left( m-n+2\right)
t+C_{2}\left( m-n+2\right) \right) ^{\frac{4+m+\omega \left( n+2\right) }{%
\left( m-n+2\right) }}}{(C_{1}t+C_{2})^{3}},
\end{equation}
and it is observed that 
\begin{equation}
\frac{G}{c^{2}}=const.\Longleftrightarrow m=\frac{3n+\omega \left(
n+2\right) }{2}-1,  \label{gc22}
\end{equation}
we are following the same argumentation as in the above model.

Once it has been obtained $G$ then we back to eq. (\ref{lam2}) obtaining in
this way the behaviour of $\Lambda :$%
\begin{equation}
\Lambda =\Lambda _{0}\left( \frac{\left( C_{1}\left( m-n+2\right)
t+C_{2}\left( m-n+2\right) \right) ^{\frac{-2n}{\left( m-n+2\right) }}}{%
\left( C_{1}t+C_{2}\right) ^{2}}-\frac{\left( C_{1}\left( m-n+2\right)
t+C_{2}\left( m-n+2\right) \right) ^{\frac{2\left( 1-n\right) +\omega (n-m)}{%
\left( m-n+2\right) }}}{\left( C_{1}t+C_{2}\right) ^{3}}\right) ,
\end{equation}
with $\Lambda _{0}=const.$

The behaviour of the Krestchmann scalars are: 
\begin{equation}
K_{1}\thickapprox \left( C_{1}t+C_{2}\right) ^{-\frac{4\left( n+1\right) }{%
m-n+2}},\text{ \ \ \ \ \ \ \ \ \ \ \ }K_{2}\thickapprox \left(
C_{1}t+C_{2}\right) ^{-\frac{4\left( n+1\right) }{m-n+2}},
\end{equation}
while the expansion and the shear behave as: 
\begin{equation}
\theta =C_{1}(m+2)\left( C_{1}t+C_{2}\right) ^{^{-\frac{m+2}{m-n+2}}},
\end{equation}
\begin{equation}
\sigma =\frac{\sqrt{3}}{3}C_{1}\left( m-1\right) \left( C_{1}t+C_{2}\right)
^{^{-\frac{m+2}{m-n+2}}}.
\end{equation}

In this case we have the following restriction:
\begin{eqnarray*}
n &<&m+2,\text{ \ \ from \ \ }Y(t), \\
3n &>&2\left( m+2\right) ,\text{ \ \ if \ }q<0, \\
n &>&-1,\text{ \ from }K_{i}, \\
m &>&-2,\text{ from }\theta \text{ and }\sigma ,
\end{eqnarray*}
which are very similar to the obtained in the above model. Now if we ``assume''
that the condition (\ref{gc22}) must be verified then it is obtained a
new restriction
\begin{equation*}
m=\frac{3n+\omega \left( n+2\right) }{2}-1.
\end{equation*}

\subsubsection{Conclusion for this model. Numerical values and graphics for
the main quantities.}

As in the above model and following the same argument, in this subsection we
will study numerically some of the possible cases in order to show the
behaviour of the different quantities. We shall study five cases, for this
purpose we need to fix some numerical constants $\left( n,m\right) $ as well
as consider different equations of state $\left( \omega \right) $. Since we
only have the above restrictions (which do not help us) we can choose
these numerical values in a very arbitrary way but we will ``assume'' that eq. (\ref
{gc22}) is verified.

In the rest of this section we shall consider the following values for the
numerical constants 
\begin{equation*}
C_{1}=1,\quad C_{2}=0,\quad c_{0}=1,\quad d_{0}=1,
\end{equation*}
and in the following table is summarized the colour of each solution which
corresponds to the different values of $\left( m,n,\omega \right) $ and the
corresponding value of the parameter $q:$ 
\begin{equation*}
\begin{array}{|c|c|c|c|}
colour & \omega  & n & m \\ \hline
red & 1 & 1/4 & 1/2 \\ \hline 
blue & 1/3 & 3/5 & 1/3 \\ \hline
magenta & 0 & 10/9 & 2/3 \\ \hline
black & -1/3 & 5/2 & 5/3 \\ \hline 
points & -2/3 & 3 & 11/6
\end{array}
\begin{array}{||c|}
q \\ \hline
17/10>0 \\ \hline
43/35>0 \\ \hline
3/4>0 \\ \hline
-0.04<0 \\ \hline
-8/23<0
\end{array}
\end{equation*}

We would like to emphasize that the black colour (line) solution i.e. it has
been calculated with the following numerical values $\left( \omega
=-1/3,n=5/2,m=5/3\right) $ does not verify the condition (\ref{gc22})$,$
the rest of the solutions verify such relationship. The only solutions that
verifies the condition $q<0$ correspond to the equation of state $\omega
=-1/3$ and  $\omega =-2/3$ which is in agreement with the recent
observations. 

\begin{figure}[h!]
\begin{center}
\includegraphics[height=2.194in,width=2.194in]{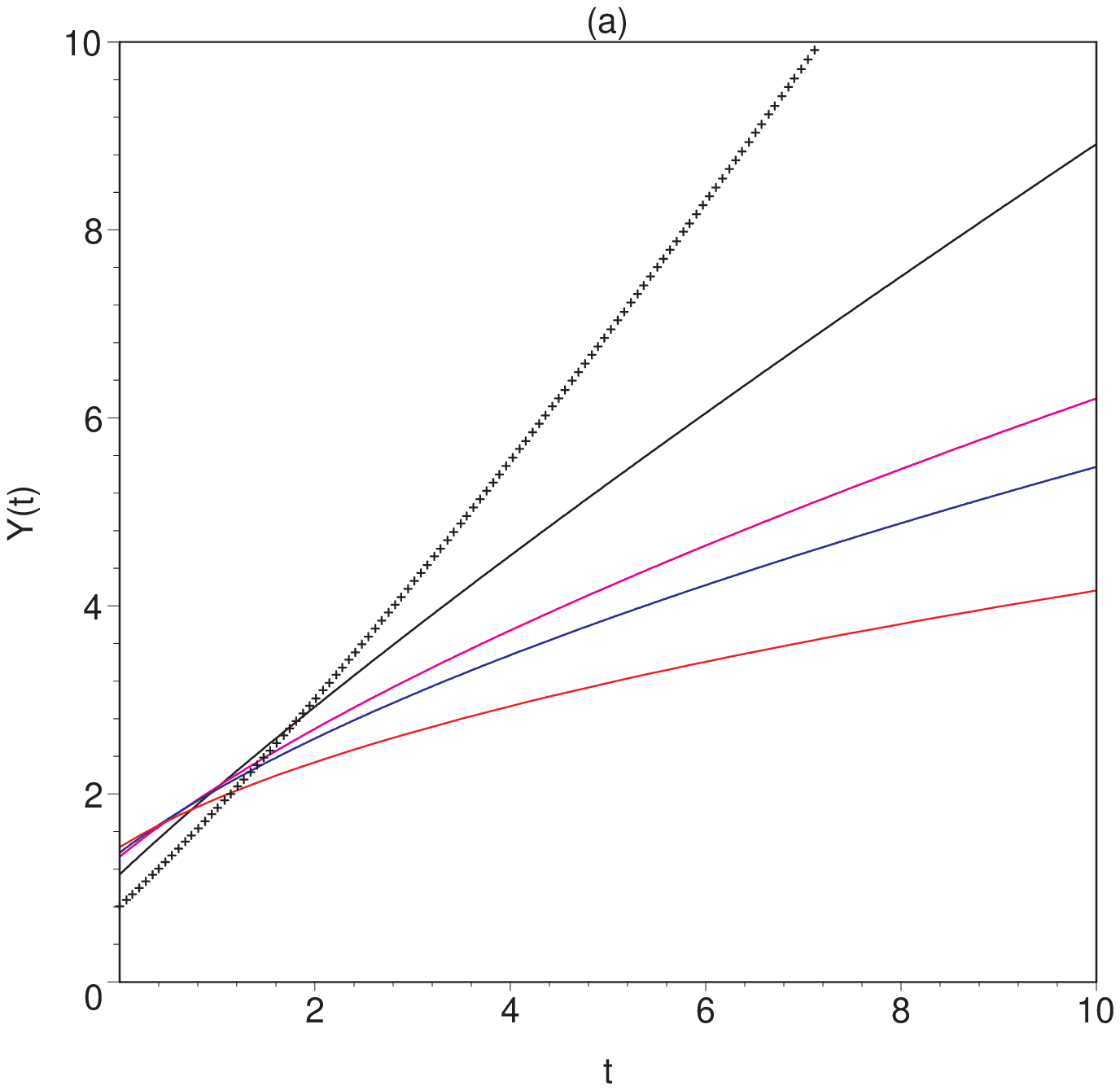} %
\includegraphics[height=2.194in,width=2.194in]{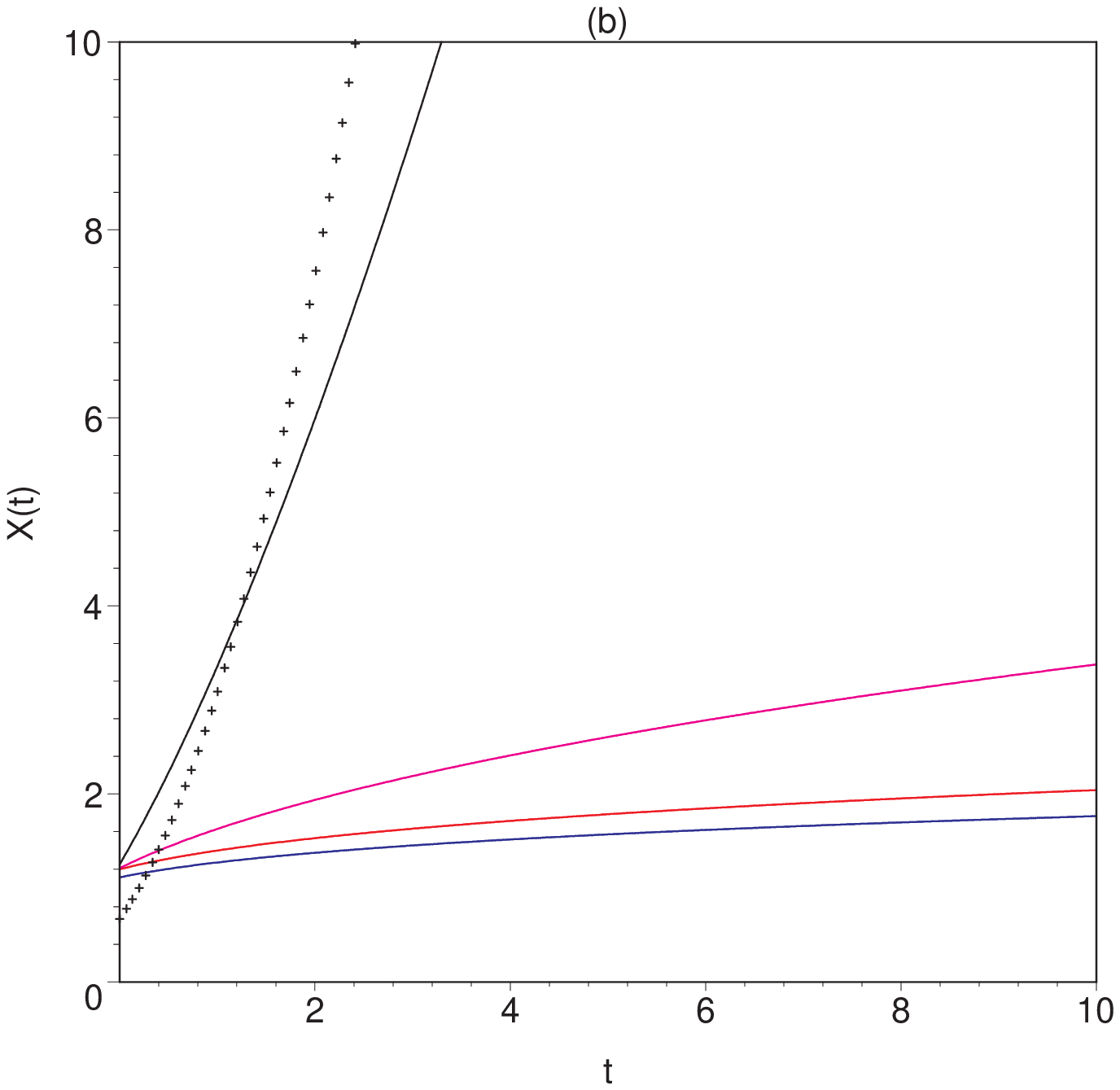}
\end{center}
\caption{In the figure (a) it is plotted the radius $Y(t)$ while in figure
(b) it is plotted the variation of the radius $X(t)$. Both radius have a
nonsingular start and are growing functions on time.}
\label{Radios3}
\end{figure}

With these numerical values we can see in fig. (\ref{Radios3}) the different
behaviour of our scale factors. In all the studied cases we can see that
these scale factors are growing functions on time $t$. These solutions are
non-singular as the Krestchmann invariants show us. 

\begin{figure}[h!]
\begin{center}
\includegraphics[height=2.194in,width=2.194in]{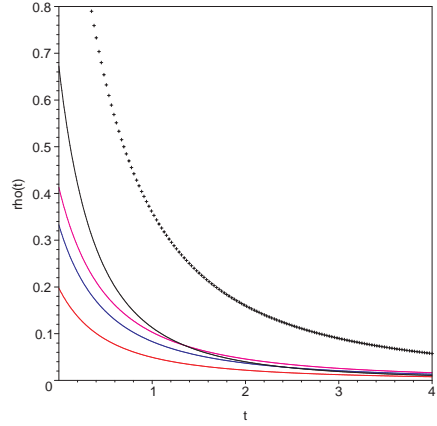}
\end{center}
\caption{Figure (a) shows the variation of the energy density $\protect\rho%
(t)$. Figure (b) shows the variation of the bulk viscous pressure $\Pi(t)$.}
\label{densidad3}
\end{figure}

With regard to the energy density (see fig. (\ref{densidad3})) we can see
that all solutions have physical meaning, since they are decreasing functions on
time $t.$

\begin{figure}[h!]
\begin{center}
\includegraphics[height=2.194in,width=2.194in]{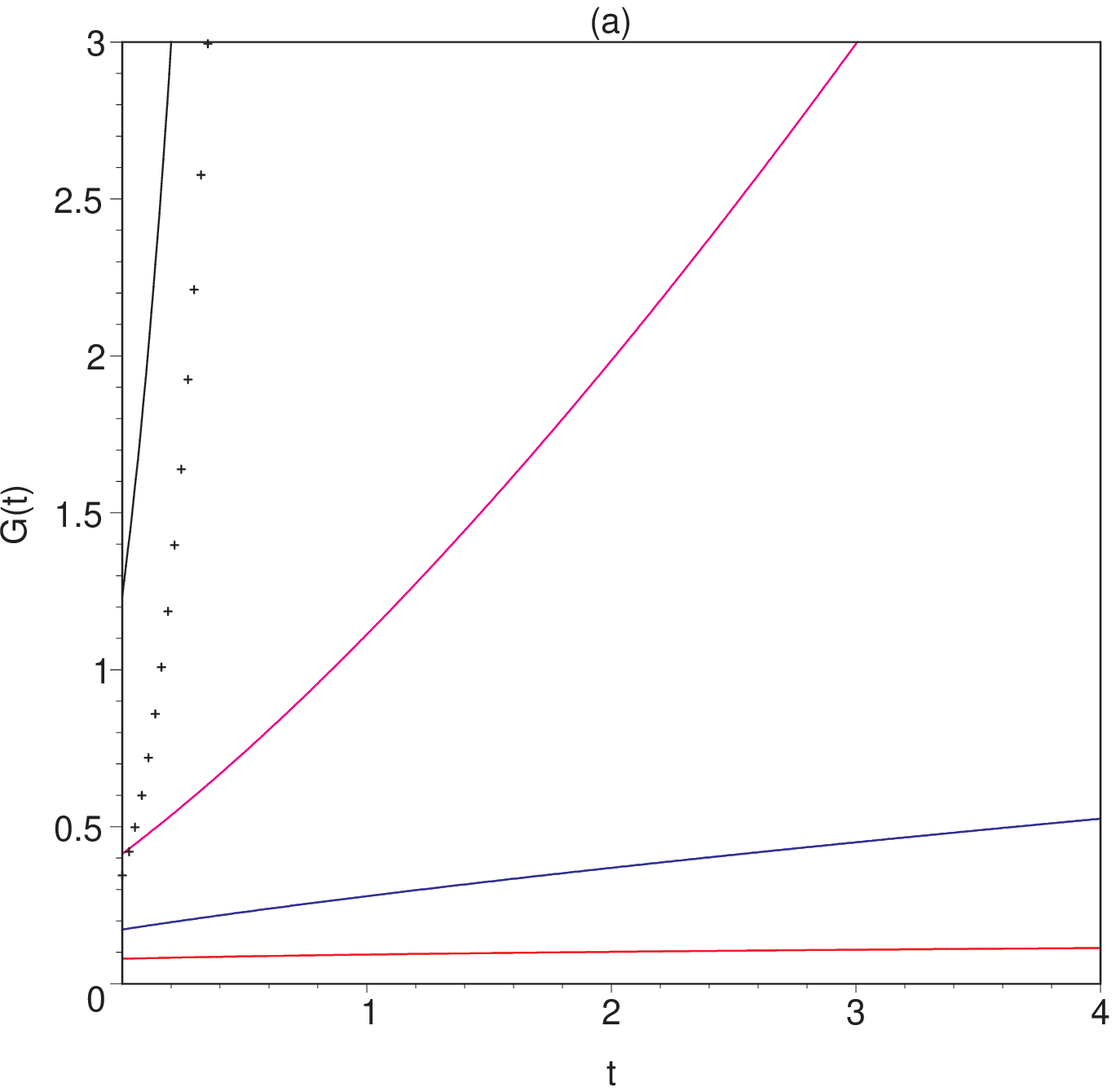} %
\includegraphics[height=2.194in,width=2.194in]{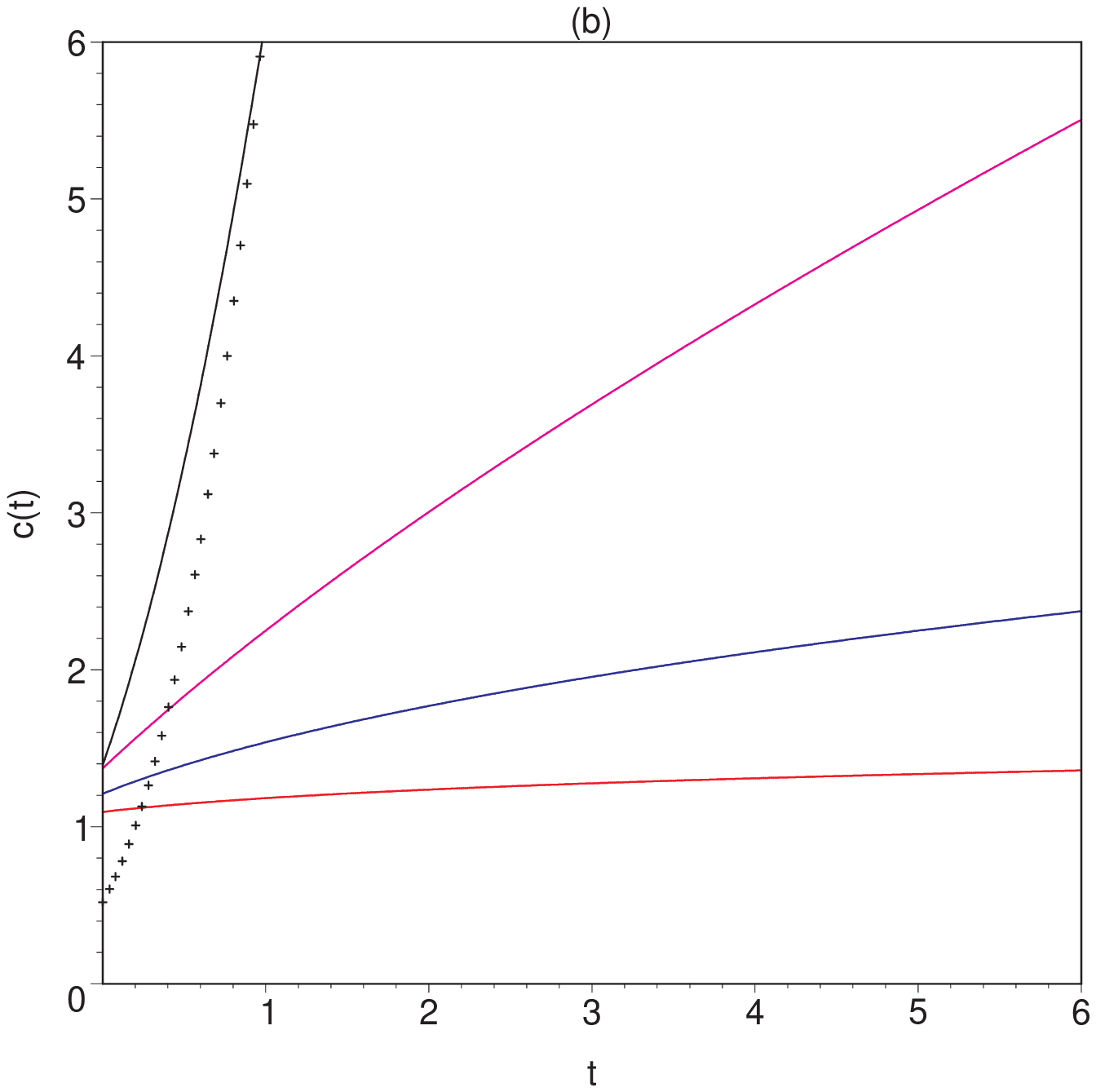} %
\includegraphics[height=2.194in,width=2.194in]{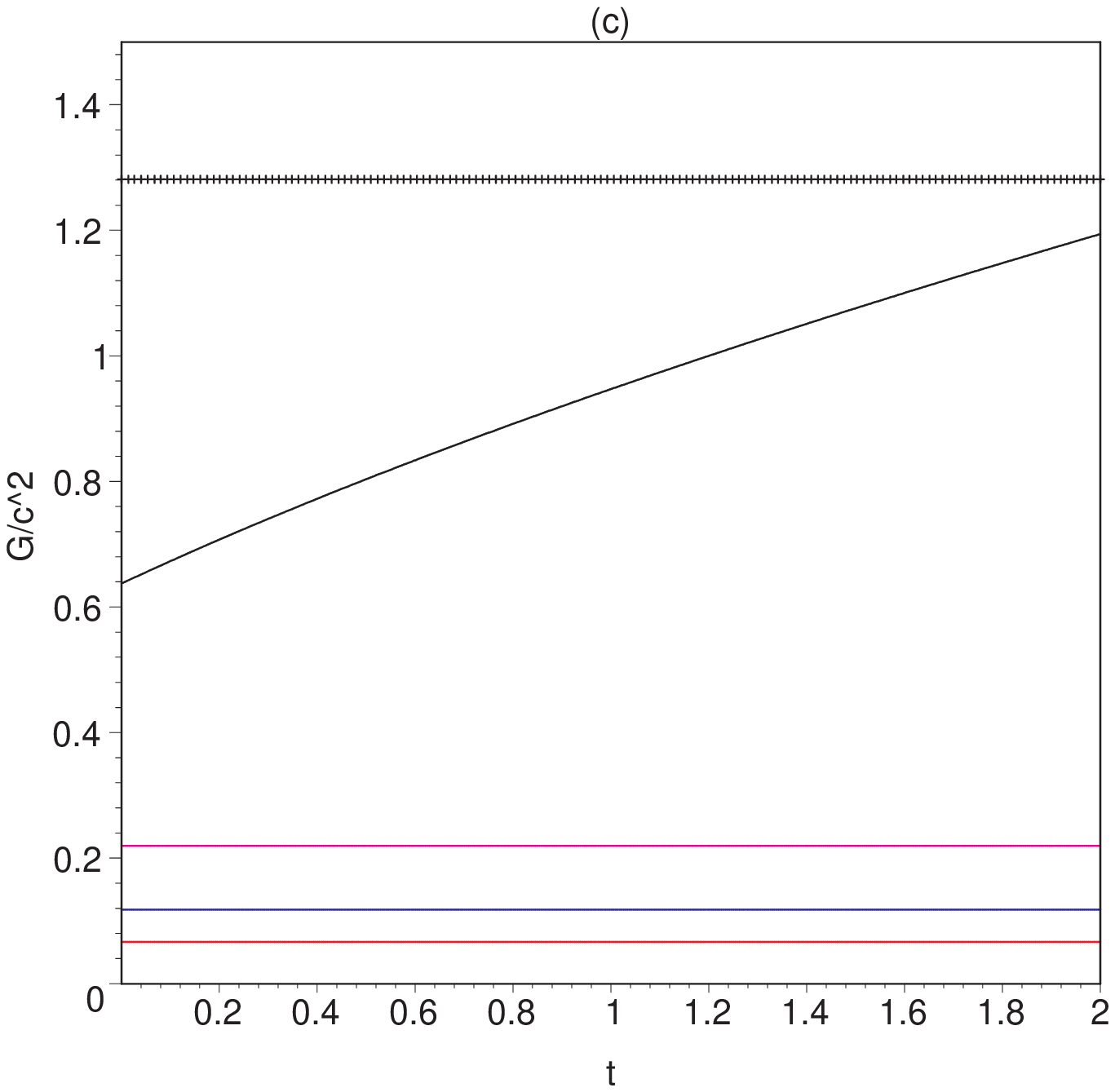}
\end{center}
\caption{In the figure (a) we show the variation of ``constant" $G(t)$. In
the figure (b) we show the variation of the ``constant" $c(t)$ and in the
figure (c) it is plotted the relationship $G/c^{2}$.}
\label{constantes3}
\end{figure}

The variation of the ``constants'' $G$ and $c$ as well as the relationship $%
G/c^{2}$ is shown in fig. (\ref{constantes3}). These pictures show us that
both ``constants'' are growing functions on time $t$. We would like to
emphasize that only for the case $\omega =-1/3$ we have that it is not
verified the relationship $G/c^{2}=const.$ as we have imposed.

\begin{figure}[h!]
\begin{center}
\includegraphics[height=2.194in,width=2.194in]{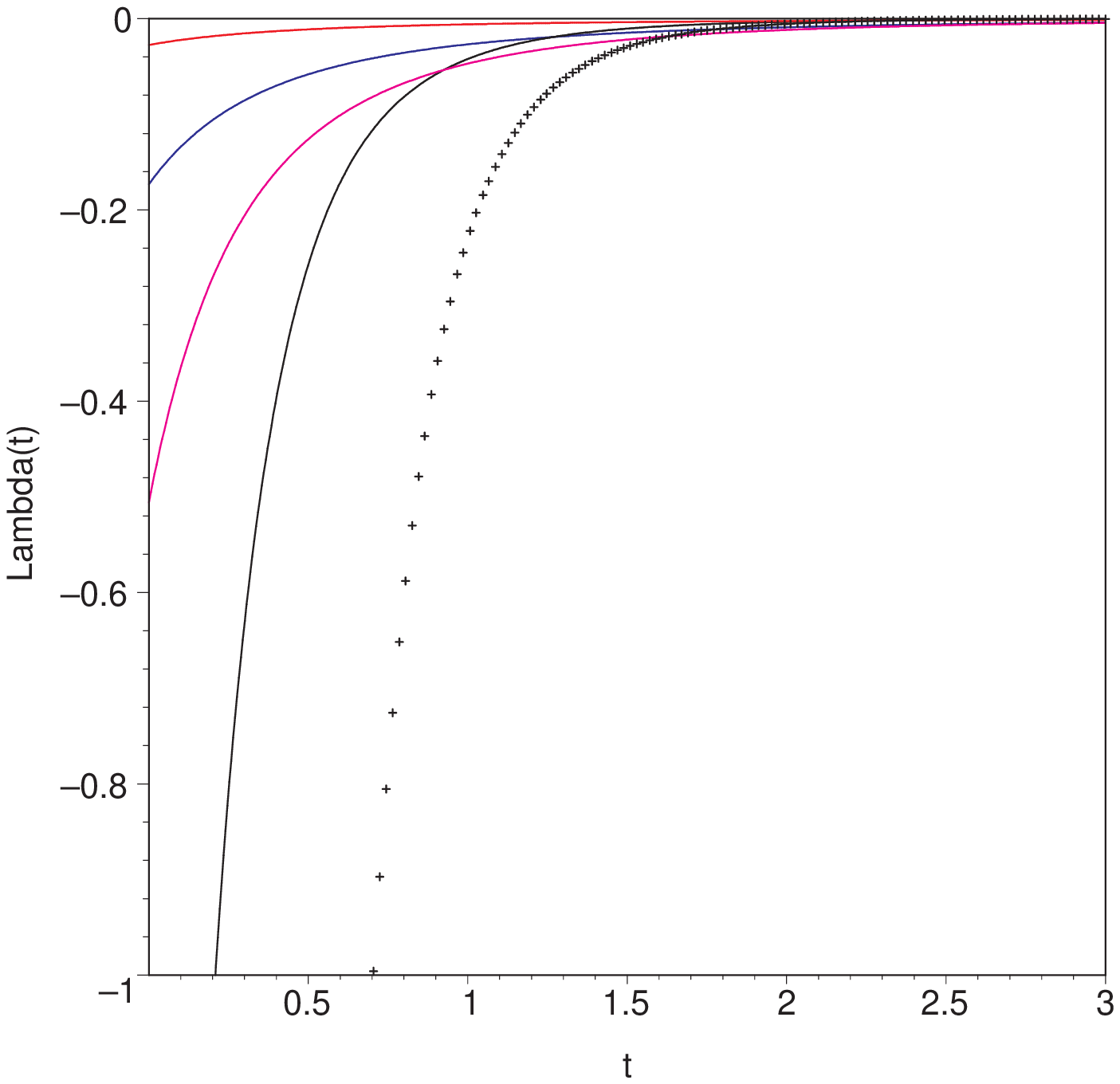}
\end{center}
\caption{The variation of the cosmological ``constant" $\Lambda(t)$.}
\label{lambda3}
\end{figure}

With regard to the cosmological ``constant'', see fig. (\ref{lambda3}), it
is observed that all the solutions are decreasing but negative. In these
cases and because our models are non-singular $\Lambda \longrightarrow \pm
const$ when $t\longrightarrow 0.$

\begin{figure}[h!]
\begin{center}
\includegraphics[height=2.194in,width=2.194in]{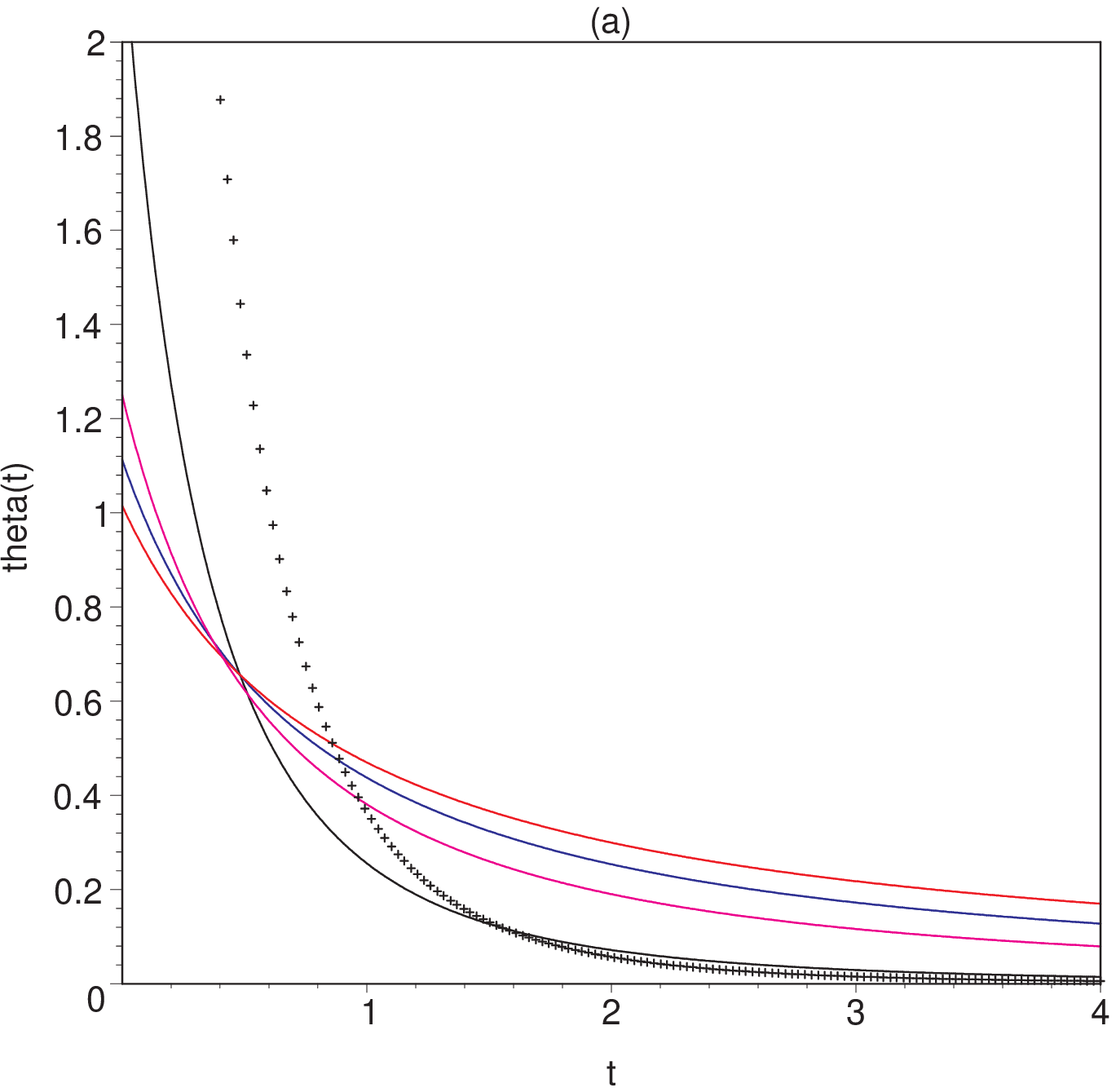} %
\includegraphics[height=2.194in,width=2.194in]{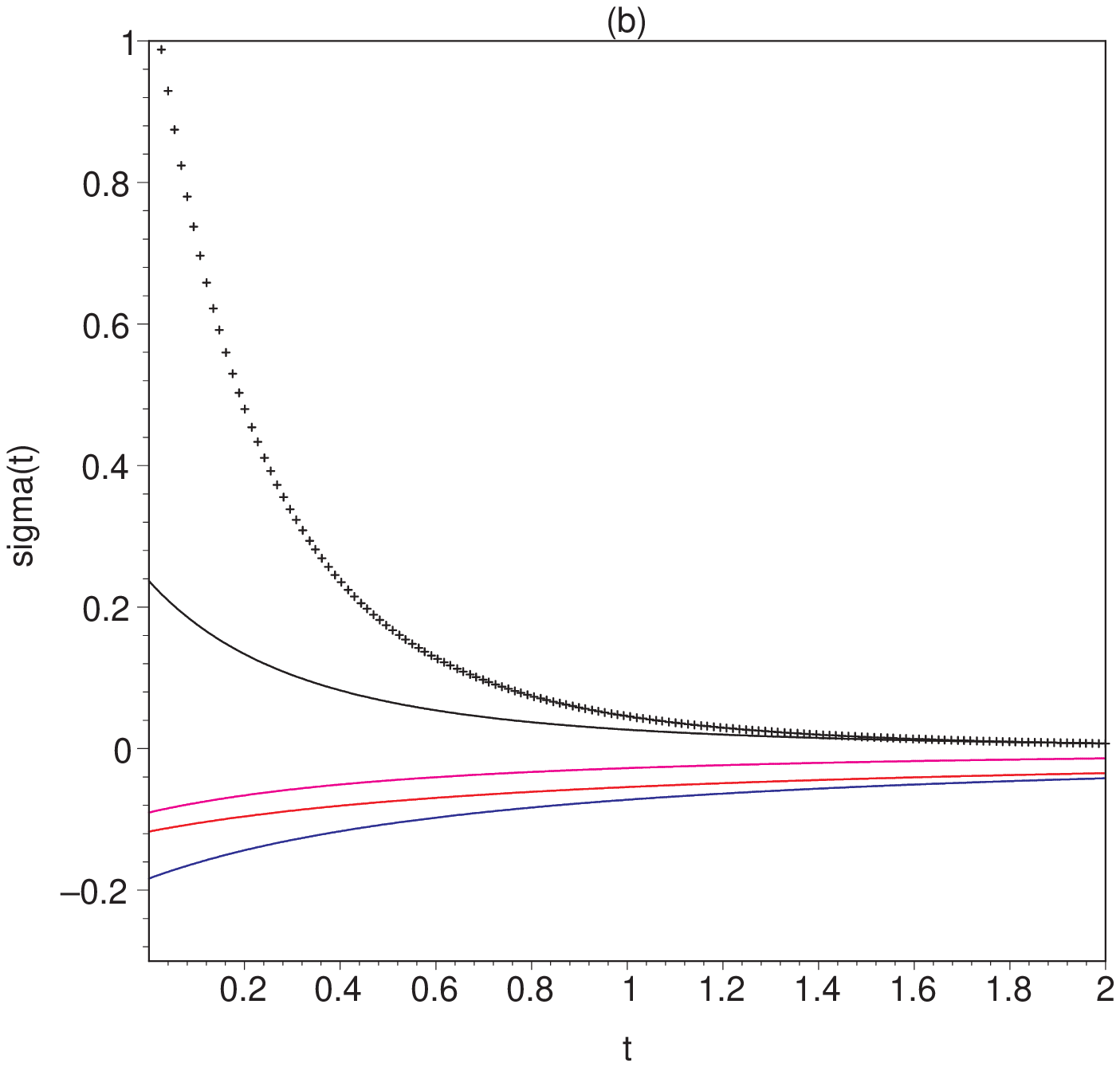}
\end{center}
\caption{In figure (a) it is plotted the expansion $\protect\theta (t)$,
while in figure (b) it is plotted the shear $\protect\sigma (t)$.}
\label{expansion3}
\end{figure}

The expansion and the shear behave as follows, see fig. (\ref{expansion3}).
As we can see all the models studied show a decreasing expansion and only
the models that have a positive shear are plotted with the red and blue
colours.

Other solutions can be studied changing the numerical values of the
constants as well as taking into account different equations of state. We
have found at least two non-singular physical solution (black line and black
points) which equations of state are $\omega =-1/3$ (strings) and $\omega
=-2/3$ (domain walls).

As we have been able to see it is very difficult to determine the behaviour
of the constants since we cannot determine or restrict better the value of
the numerical constants that define such behaviour. While in the viscous
case we arrive to the conclusion that $G$ and $c$ only could behave as
growing functions on time $t$, in these models we have founded that a
decreasing behaviour is allowed as well as the growing one. Only if we
insist in that our model verifies the condition $q<0$ we find that $G$ and $c
$ must be growing functions while $\Lambda $ is a negative decreasing
function on time $t.$

We have chosen our numerical values under the ``assumption'' that $%
G/c^{2}=const.$ (fine tuning) but of course this restriction is arbitrary
although in the viscous model was verified (only for $\gamma =1/2,$ fine
tuning again). With out other physical considerations we cannot ensure that $%
G$ and $c$ are growing or decreasing functions on time, nevertheless, it
would seem reasonable to choose growing functions in order to connect the
viscous era to the perfect fluid era. In both cases $\Lambda $ is a negative
decreasing function.

\vspace{0.5cm} \textbf{Acknowledgment} I wish to acknowledge to Javier
Aceves his translation into English of this paper.


\begin{thebibliography}{9}

\bibitem{Tony2}  {\footnotesize J.A. Belinch\'{o}n. ``Bulk Viscous LRS
Bianchi I with Time-varying Constants''. gr-qc/0410065}



\bibitem{Ha}  {\footnotesize J. Hajj-Botros and J. Sfeila. \textit{Inter.
Jour. Theor. Phys.} \textbf{26},97, (1987) }

\bibitem{Ram}  {\footnotesize S. Ram. \textit{Inter. Jour. Theor. Phys.} 
\textbf{28},917, (1989) }

\bibitem{M}  {\footnotesize A. Mazumber. \textit{Gen. Rel. Grav.} \textbf{26}%
,307, (1994) }

\bibitem{vis}  {\footnotesize R.G. Vishwakarma and Abdusssatar. \textit{%
Phys. Rev. D} \textbf{60}, 063507, (1999) }

\bibitem{Cha}  {\footnotesize I. Chakrabarty and A. Pradhan. \textit{Grav.
and Cos.} \textbf{7}, 55, (2001).}




\bibitem{Tony1}  {\footnotesize J.A. Belinch\'{o}n and J.L. Caram\'{e}s. ``A
New Formulation of a Naive Theory of Time-varying Constants.'' gr-qc/0407068 
}

{\footnotesize \ }
\end{thebibliography}
\end{document}